\documentclass[doublecol]{epl2}

\usepackage{units}
\usepackage{graphicx}
\usepackage{graphics}
\usepackage{epstopdf}
\usepackage{subfig}

\title{Defect-induced phase transition in the asymmetric simple exclusion process}
\shorttitle{Defect-induced phase transition in the TASEP}

\author{Johannes Schmidt\inst{1} \and Vladislav Popkov\inst{1,2}
\and Andreas Schadschneider\inst{1}}
\shortauthor{J. Schmidt  \and V. Popkov \and A. Schadschneider}

\institute{
  \inst{1} Institut f\"ur Theoretische Physik, Universit\"at zu K\"oln,
50937 K\"oln, Germany\\
  \inst{2} CSDC Universit\`a di Firenze, via G.Sansone 1,
50019 Sesto Fiorentino, Italy
}

\abstract{
We reconsider the long-standing question of the critical
  defect hopping rate $r_c$ in the one-dimensional totally asymmetric
  exclusion process (TASEP) with a slow bond (defect). For $r< r_c$
  a phase separated state is observed due to queuing at the defect site
  whereas for $r\geq r_c$ the defect site has only local effects on
  the stationary state of the homogeneous system.  Mean-field theory
  predicts $r_c=1$ (when hopping rates outside the defect bond are equal to 1)
  but numerical investigations seem to indicate $r_c \approx 0.80(2)$. Here
  we improve the numerics to show that $r_c > 0.99$ and give strong evidence that indeed
  $r_c=1$ as predicted by mean-field theory, and anticipated by recent theoretical findings.
  }

\begin{document}

\maketitle


\section{Introduction}

Despite much progress in recent years, our understanding of
nonequilibrium stationary states is far from complete. This
especially concerns the effects of disorder and defects in
driven diffusive systems. Although it is well established
that in driven systems already a localized defect can have
a global influence on the behavior, many open questions remain.
Since e.g. no analogue of the Harris criterion \cite{Harris}
is known for nonequilibrium situations no general statements
on the influence of weak disorder on the critical behavior can
be made \cite{Stinchcombe}.

Surprisingly even for the simplest model of driven diffusion,
the totally asymmetric exclusion process (TASEP), the precise
influence of a single defect has not been fully clarified for a long time.
It is well-known since the seminal work of Janowsky and Lebowitz
\cite{JanLeb1,JanLeb2} that such a defect has not just a local
effect, but changes the nature of the stationary state dramatically
(for reviews, see e.g.\ \cite{Krug,Barma}).
What is not clear up to now is whether a finite critical strength of the
defect is required to create global effects. Mean-field theory predicts
a global influence already for arbitrarily small defect strengths
whereas the most accurate numerical investigations up to date
\cite{Meesoon_ERROR} show strong indications that a finite defect
strength is required.

Recently the problem has been newly addressed in the mathematical literature, in a series
of works \cite{Costin_Lebo,Sidoravicius2015,Calder2015,Scoppola2015}.
In \cite{Costin_Lebo}, based on analytical arguments from series expansions and
results for related systems (e.g. directed polymers) it was argued that
an arbitrarily small defect in a TASEP with open boundaries will have global effects, e.g. on the current
and on the density profile.
In \cite{Sidoravicius2015}, the authors claim to have proved rigorously,
that the steady current in TASEP with a slow bond is always affected for any nonzero defect strength.

In view of these new findings, the numerical studies predicting finite critical blockage strength,
 appear even more  paradoxical, with the $r_c=1$ problem finally settled.  It remains to understand if the effects of the slow bond are so weak that they cannot be
observed in numerics, which would make the beautiful theoretical result a pure theoretical construction
without any practical content.

It is the purpose of this letter to show that this is not the case, and
to provide detailed results from highly accurate Monte Carlo
simulations which strongly support the theory. Moreover, we also resolve the apparent numerical paradox,
revisiting previous numerical studies and pointing out exactly where the error of the previous numerical studies was.


\section{Model}

The totally asymmetric simple exclusion process (TASEP) is a
paradigmatic model of nonequilibrium physics (for reviews, see e.g.
\cite{bib:L,bib:Schue,bib:ZS,bib:SCN,bib:KRB,bib:BE,bib:D}) describes
interacting (biased) random walks on a discrete lattice of $N$ sites,
where an exclusion rule forbids occupation of a site by more than one
particle. A particle at site $k$ moves to site $k+1$ at rate $p$ if
site $k+1$ is not occupied by another particle. The boundary sites
$k=1$ and $k=N$ are coupled to particle reservoirs.  If site $1$ is
empty, a particle is inserted at rate $\alpha$.  A particle on site
$N$ is removed from the system at rate $\beta$. Sites are updated
using random-sequential dynamics. Throughout the paper, we will set
$p=1$.

\begin{figure}
\begin{center}
\includegraphics[width=8cm]{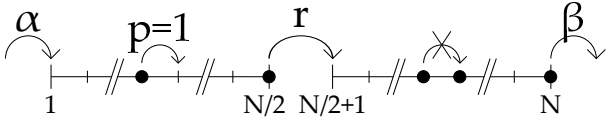}
\par\end{center}
\caption{Open TASEP with a slow bond ($r<1$) in the middle.}
\label{Fig_Defect_TASEP_Model}
\end{figure}

Here we consider a system of two TASEPs of length $N/2$ coupled by
a slow bond between sites $N/2$ and $N/2+1$ with reduced hopping
rate $r\leq p$ (Fig.~\ref{Fig_Defect_TASEP_Model}). This is equivalent
to a TASEP of $N$ sites with a defect site in the middle and defect
strength $p-r$.

For periodic boundary conditions this problem has been analyzed
by Janowsky and Lebowitz \cite{JanLeb1,JanLeb2}.
Below a critical rate $r_c$ they found a phase separation into high
and low density regions due to queuing at the defect site.
The two phases are separated by a shock (domain wall).
The phase separation is also reflected in the current-density
relation (fundamental diagram) which shows a density-independent
current at intermediate densities due to the current-limiting effect
of the slow bond.  A mean-field theory that neglects correlations at
the defect site predicts that $r_c=1$ \cite{JanLeb1}, i.e.\
an arbitrarily small defect leads to a phase separated stationary state.
This is supported by series expansions obtained from exact results
for small systems \cite{JanLeb2}.
Exact results have been obtained for the case of sublattice-parallel
update with deterministic bulk hopping by Bethe Ansatz \cite{Schuetz1}
and matrix-product Ansatz \cite{Hinrichsen}. Also in this case
$r_c=1$.

For open boundary conditions, a mean-field treatment of the TASEP with
a defect in the middle of the system yields $r_c=1$
\cite{kolomeiski2}.  Later Ha et al.~\cite{Meesoon_ERROR} studied the
problem numerically (with rates $p=1$, $\alpha=\beta=1/2$).  They
suggested that $r_c=0.80\left(2\right)$, see next section for details.

Due to its relevance e.g. for intracellular transport, recently the
open TASEP with several defects has attracted some attention, see e.g.
\cite{barma2,chou,LakatosBC,harris1,Enaud,juhasz,frey2,dong,foulaad,GreulichS1,GreulichS2,GreulichS3}.

\section{Simulations}
\label{sec-rc}

In order to analyze the system with a defect we perform Monte Carlo (MC)
simulations for an open TASEP on a chain of large size ($N\leq
200.000$), to minimize finite-size effects as much as possible.
Both bulk hopping rates and boundary hopping rates are chosen equal to
$1$. This corresponds to a point in the maximal current phase which is
characterized by a spatially homogeneous steady state with bulk
density of particles $\rho_{\rm bulk}=1/2$ and the current
$j(\infty)=\rho_{\rm bulk}(1-\rho_{\rm bulk})=1/4$ in the limit of an
infinite system size \cite{bib:Schue}.

Initially $N/2$ particles are distributed randomly within a
homogeneous chain without a defect.
Then, the relaxation of the system is performed for $100 N^2$
single Monte Carlo updates \cite{MCupdate},
 according to the dynamical rules
for $\alpha=\beta=1$.

\begin{figure}
\begin{center}
\includegraphics[width=8cm]{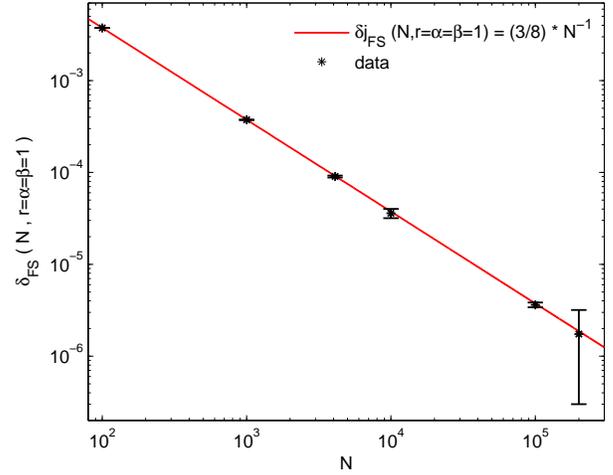}
\par\end{center}
\caption{Finite-size corrections of the steady current for a
  homogeneous model without slow bond. Error bars show the
  99\% confidence bound, the red line marks the exact
  leading finite-size correction in $1/N$.}
\label{Fig_FSS_r1}
\end{figure}

After the initial relaxation, the weak bond is introduced in the
middle of the system, meaning that a particle hop from site $k=N/2$ to
$N/2+1$ occurs with reduced rate $r<1$, see
Fig.~\ref{Fig_Defect_TASEP_Model}.  Then the system is relaxed further
for $100 N^2$ single Monte Carlo updates
and the average current is recorded. It can be written as
\begin{equation}
j_{\rm FS}(r,N,\alpha,\beta)= j(r,\infty) + \delta _{\rm FS}(r,N,\alpha,\beta)
\label{MC_CURRENT}
\end{equation}
where $\delta_{\rm FS}$ are finite-size corrections.
The relaxation to steady state is controlled by a comparison of the
finite-size corrections of the current measured numerically with the
theoretically-predicted value, see Fig.~\ref{Fig_FSS_r1}.

\begin{figure}
\begin{center}
\includegraphics[width=8cm]{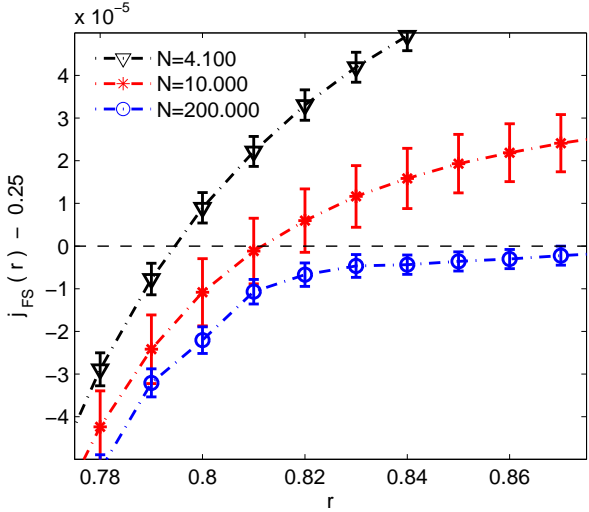}
\par\end{center}
\caption{Finite-size current $j_{\rm FS}$ (see eq.~(\ref{MC_CURRENT})) versus
  defect rate $r$ for $\alpha=\beta=1$. With $j\left(r\right)\leq
  j_{FS}\left(r\right)$ the plot shows $r_c>0.86$ which is significantly
  larger than $r_c=0.80\left(2\right)$ suggested in
  \cite{Meesoon_ERROR}. Currents are averaged over all sites with
  $7.5\cdot10^7-5\cdot10^8$ histories. Error bars show the 99\%
  confidence interval.}
\label{Fig_current_Zoom}
\end{figure}

We aim at determining a lower bound for the critical $r_c$ at which the
phase separation starts. It is well-known \cite{DerridaEvans93}
that the leading finite-size corrections to the current of the
  homogeneous TASEP for $\alpha=\beta=1$ are positive,
\begin{eqnarray}
  j_{\rm FS}(r=1,N,\alpha=1,\beta=1) &=&
  j(r=1,\infty)+\frac{3}{8N}+ O(N^2)\nonumber\\
  &>& j(r=1,\infty)\,.
\end{eqnarray}
Therefore, if for some defect hopping rate $r_0$ the steady current
within the error bars is \textit{smaller} than its limiting value,
$j(r_0,N)<1/4$, this would definitely mean that $r_c>r_0$. The lower
bound $r^*$ for $r_c$ is then calculated as the point where
$j(r^*,N)=1/4$, accounting also for error bars, see
Fig.~\ref{Fig_current_Zoom}.  Note that
our reasoning does not involve any a priori assumptions except from
the positiveness of finite-size corrections to the current.  In this
way, we obtain a lower bound for $r_c$ which depends on the system
size. For larger system size, finite-size corrections are smaller and
better estimates for the lower bound can be made, see
Fig.~\ref{Fig_current_Zoom}.  For system size $N=2\cdot 10^5$ we
obtain the following lower bound estimate,
\begin{equation}
r_c >0.86\,.
\label{0.86}
\end{equation}
It is difficult to improve the lower bound (\ref{0.86}) by a further
increase of the system size $N$, because much larger system sizes are
not numerically accessible. However, already the lower bound value
$0.86$ definitely contradicts the result of Ha et
al.~\cite{Meesoon_ERROR}, where $r_c=0.80\left(2\right)$ was found.
Their estimate was based on a different argument. In order
to understand the reason for the contradiction, we repeated Monte
Carlo simulations with the same parameters as in \cite{Meesoon_ERROR},
and in particular using the much smaller system size of $N=4100$.

\begin{figure}
\begin{center}
\includegraphics[width=8cm]{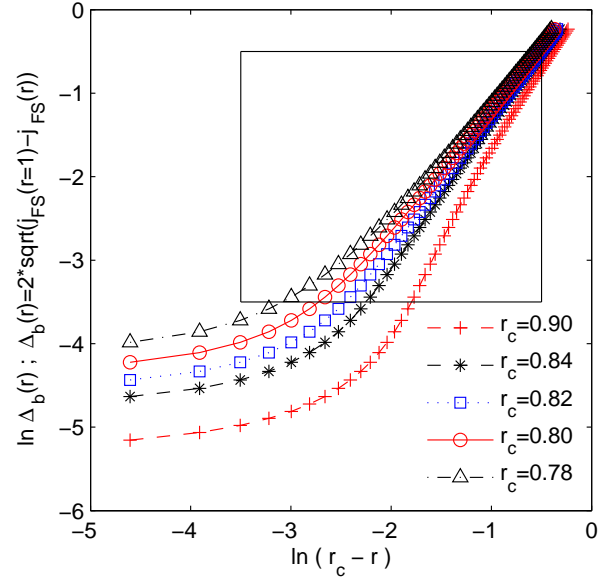}
\end{center}
\caption{Double logarithmic plot of $\Delta_b$ versus $\left(r_c-r\right)$.
  Parameters are $N=4100,\alpha=\beta=0.5$.  The box indicates the
  regime plotted in Fig.~4a of \cite{Meesoon_ERROR}.  Averaging over
  all sites of the system and over $3\cdot10^8$ histories is
  performed.  Statistical errors are smaller than the symbol size.}
\label{Meesoon_delta_b_scaling}
\end{figure}

The key quantity analyzed by Ha et al., is defined as
\begin{equation}
\label{def_delta_b}
\Delta_b \propto \left| \rho_{\rm bulk,segment}-\frac{1}{2} \right|\,
\end{equation}
or
\begin{equation}
\Delta_b =2 \sqrt{j(r=1,N)-j(r,N)} \,.
\label{delta_b_current}
\end{equation}
It is assumed to obey the scaling form
\begin{equation}
\Delta_b \sim (r_c-r)^{-\beta} \label{delta_b_scaling}
\end{equation}
near the phase boundary. Then, the best straight line fit on the
double logarithmic plot of $\Delta_b$ versus $r_c- r$, has lead
to the conclusion $r_c=0.80\left(2\right)$.  We repeated the relevant
Monte Carlo simulations for the parameters chosen in
\cite{Meesoon_ERROR}. Fig.~\ref{Meesoon_delta_b_scaling} shows
the double logarithmic plot of $\Delta_b$ versus $r_c- r$.
The square window corresponds to the area shown in the original paper,
see Fig.~4a of \cite{Meesoon_ERROR}. We can see that what might look as
a straight line inside the window, certainly fails to straighten
outside the window. Thus the conclusion of \cite{Meesoon_ERROR} of a
phase transition at $r_c=0.80\left(2\right)$ is not justified.

\begin{figure}
\begin{center}
\includegraphics[width=8cm]{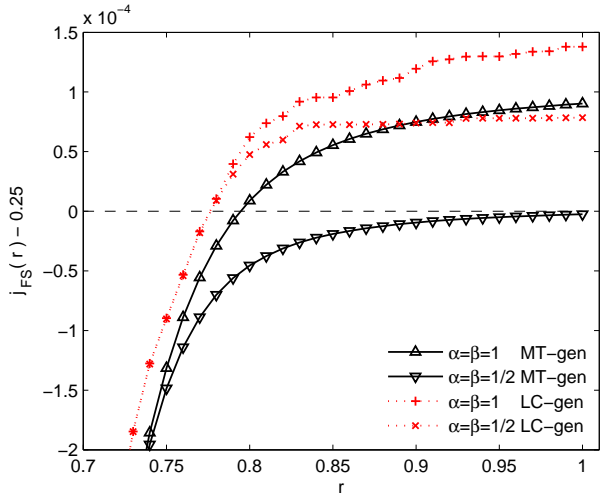}
\par\end{center}
\caption{Finite-size corrections of the current eq.~(\ref{MC_CURRENT}),
  for a different choices of a random number generators.  Parameters:
  $N=4100$. Triangles correspond to Mersenne Twister random number
  generator and crosses correspond to a standard Park and Miller new
  minimal standard (LC-gen, linear congruential) random number
  generator \cite{Park_Miller}.  Averaging of the current over all
  bonds and over $3\cdot10^8-5\cdot10^8$ histories is performed.
  Statistical errors are smaller than the symbol size.}
\label{CURRENT_GENERATOR}
\end{figure}

It is instructive at this point to stress the importance of a choice
of an adequate random generator to perform the Monte Carlo update.
This choice is crucial for producing high quality Monte Carlo data
\cite{Random_TEST}.  In Fig.~\ref{CURRENT_GENERATOR}, Monte Carlo data
for the current, produced by different random number generators, are
compared and show systematic differences. For our simulations
throughout this paper we are using Mersenne-Twister-generator which is
known for producing high quality pseudo-random numbers. In
Fig.~\ref{CURRENT_GENERATOR} it is seen that using the most common
Park and Miller new minimal standard linear congruential generator
\cite{Park_Miller} leads to a systematic overestimation of the
current, and might consequently lead to wrong conclusions in the
subtle TASEP blockage problem.

We also note, that the point $\alpha=\beta=1/2$ as chosen for studying
the blockage problem in \cite{Meesoon_ERROR} lies exactly at the phase
boundary between the maximal current, low density and high density
phases \cite{bib:L,bib:Schue,bib:ZS,bib:SCN,bib:KRB,bib:BE,bib:D}.
This may lead to further complications and an additional reduction of
the steady current due to fluctuations.  We stress that for our study
we choose $\alpha=\beta=1$, i.e.\ a point well inside the maximal
current phase far from the boundaries with the low density TASEP phase
($\alpha=1/2$) and the high density TASEP phase ($\beta=1/2$).


\section{Effects of a defect in finite systems: parallel evolution}

As is already mentioned, both mean-field theory and series expansions
arguments hint at $r_c=1$.
Assuming the existence of an essential singularity at $r_c=1$, $j(1)-j(r)\sim \exp(-a/(1-r))$ \cite{Costin_Lebo},
further
improvement of the lower bound for the weak bond problem by an increase
of the system size is a hopeless enterprise: e.g. a numerical proof
$r_c>0.9,r_c>0.95,r_c>0.99$) with the direct method (see
Fig.~\ref{Fig_current_Zoom}) would require $N>10^{10}$, $N>10^{22}$,
$N>10^{147}$ respectively.

Instead of increasing the system size, we address the problem of a
critical blockage strength in different way by measuring how a TASEP
responds to a slow bond, as discussed below.

After the pre-relaxation performed on the homogeneous system as
described above, we make two copies of the system configuration. Then
a slow bond is introduced in one copy whereas the other remains
homogeneous. Both copies evolve in time according to the same
protocol, i.e. using the same set of random numbers for both systems.
The averaged density profiles of both copies are compared after
sufficiently large relaxation time.  Due to the defect site, one
expects a density gradient forming locally in the vicinity of the
blockage for any $r<1$. If the disturbance remains local, the
state of the system far from the blockage will not change, with
respect to a homogeneous system.  In contrast, a non-local disturbance
spreading to the whole system would lead to a reduction of the global
current and to phase separation. Thus the current and the density
profile are sensitive probes for the effects of the slow bond and for
the possible occurrence of a phase separated state.

\begin{figure}
\begin{center}
\includegraphics[width=8cm]{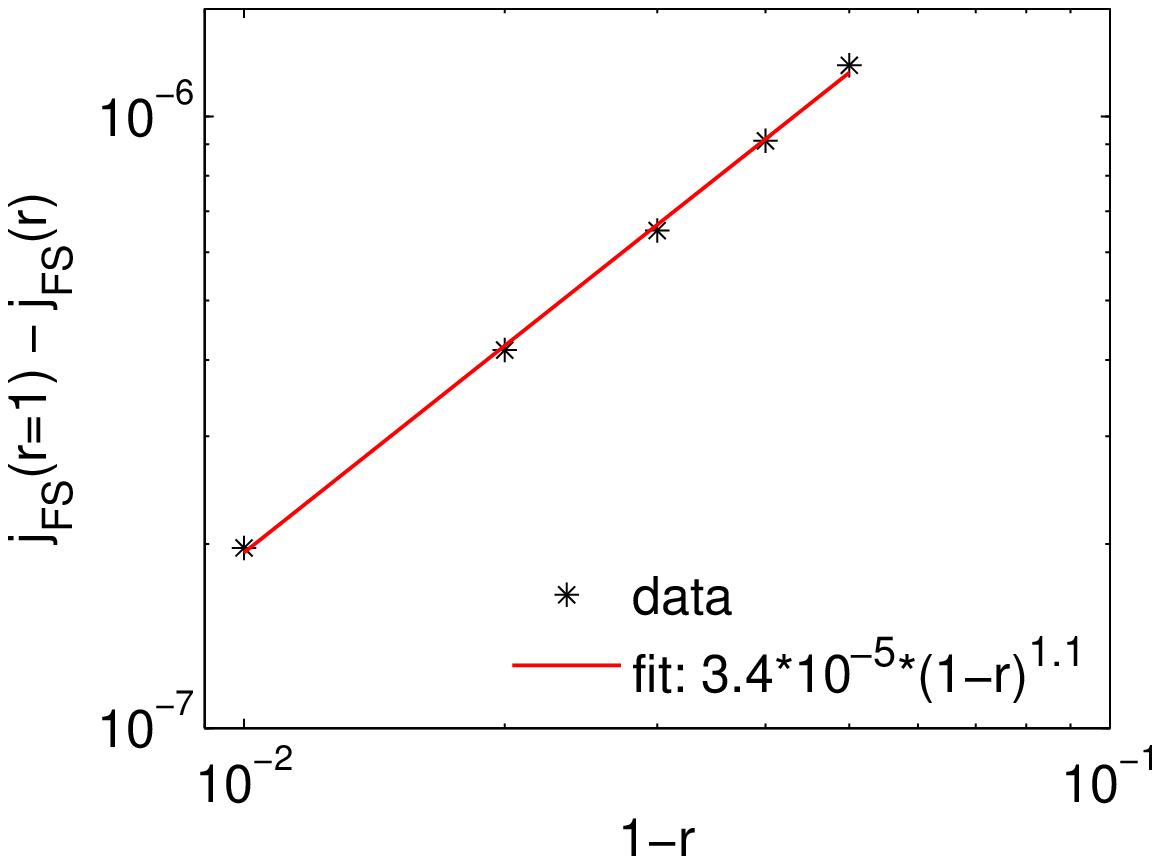}
\includegraphics[width=8cm]{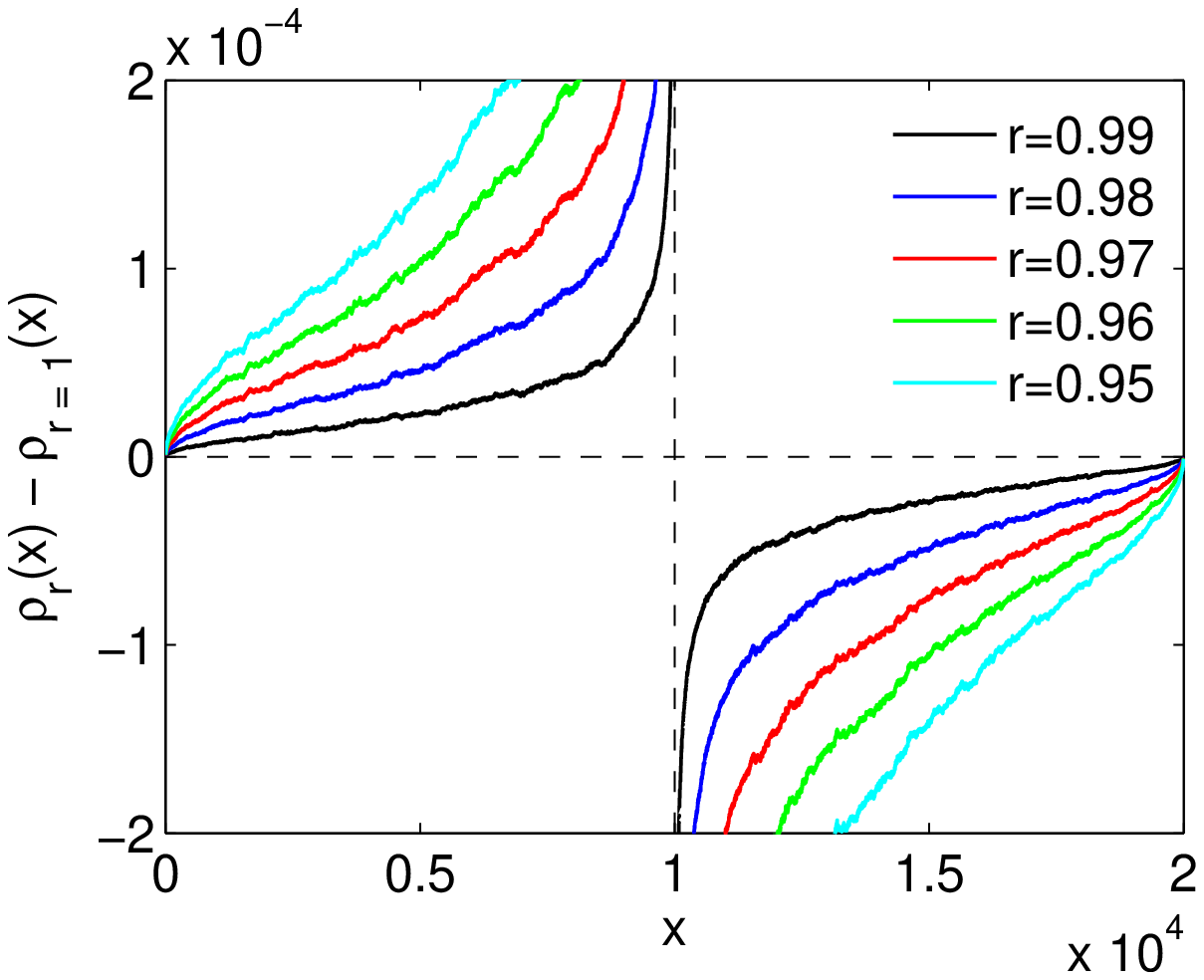}
\end{center}
\caption{Steady current differences (top) and density
  differences (bottom) between the parallel evolving copies of the
  same system (with and without defect).
  Averaging over $10^{9}$ histories is performed for systems with
  $N=2 \cdot 10^4$ sites.
  For the currents (top) error bars (not shown) are of the order of the
  symbol size.  For the density differences (bottom), statistical errors
  are in the range from $10^{-8}$ close to the reservoirs up to
  $4.6\cdot 10^{-6}$ in the vicinity of the defect.
  }
\label{fig-curr_dens_DIFF}
\end{figure}

Performing extensive MC simulations we are able to see a non-local
effect of the blockage up to $r=0.99$, both in steady current and in
local particle density far away from the blockage, see
Fig.~\ref{fig-curr_dens_DIFF}.  Consequently, a presence of the weak
bond has a small but systematic effect on both the current and the
local density $n_k$ far away from the blockage site. This shows that
the blockage produces perturbations which do not remain local, but
spread over the bulk.

\section{Conclusion}

We have revisited a long-standing problem of a TASEP with a weak bond.
Since the effects are small for small defect strength, this is a
subtle problem that requires high numerical accuracy in simulations.
We have shown that here even the choice of the random number generator
is crucial.  By Monte Carlo simulations performed on large systems (up
to $200.000$ sites), we have established a new lower bound for the
critical defect strength leading to phase separation.
For the current, we could clearly show a reduction compared
to the value $j(r=1)=1/4$ of the homogeneous system for defect
hopping rates up to $r=0.86$. Therefore we conclude that $r_c>0.86$.
Studying a parallel evolution of two initially identical systems, with
and without a weak bond, we find systematic {\em global} effects
induced by the weak bond for defect hopping rate up to $r\leq 0.99$.
Our study supports the hypothesis that the critical blockage hopping rate
is $r_c=1$, and definitely rules out a previously obtained critical
value $r_{c}=0.80(2)$. This indicates that the mean-field theory
prediction is indeed correct which is important since it is
used quite frequently also in more complex situations, like
flows on networks \cite{Embley,Ezaki,Raguin,Neri,Meesoon-recent}
or as effective models for highway traffic near ramps \cite{Diedrich}.



\acknowledgments
We thank Meeson Ha for providing background information on their work
\cite{Meesoon_ERROR} and Gunter M.~Sch\"utz for discussions.
Financial support by the DFG under grant SCHA 636/8-1 is gratefully
acknowdledged.


\begin{thebibliography}{99}

\bibitem{Harris}  A.B. Harris: J. Phys. C 7, 1671 (1974)

\bibitem{Stinchcombe} R. Stinchcombe: J. Phys.: Condens. Matter 14, 1473 (2002)

\bibitem{JanLeb1}
S.A. Janowsky, J.L. Lebowitz: Phys. Rev. A 45, 618 (1992)

\bibitem{JanLeb2}
S.A. Janowsky, J.L. Lebowitz: J. Stat. Phys. 77, 35 (1994)

\bibitem{Krug}
J. Krug: Braz. J. Phys. 30, 97 (2000)

\bibitem{Barma}
M. Barma: Physica A 372, 22 (2006)

\bibitem{Meesoon_ERROR}
M. Ha, J. Timonen, M. den Nijs: Phys. Rev. E 58, 056122 (2003)


\bibitem{Costin_Lebo}
O. Costin, J. L. Lebowitz, E. R. Speer, A. Troiani: Bulletin of the
Institute of Mathematics Academia Sinica 8, 49 (2013)

 \bibitem{Sidoravicius2015}R. Basu, V. Sidoravicius and A. Sly,
arXiv:1408.3464

\bibitem{Calder2015}  J.Calder, J. Stat. Phys. \textbf{158}, 903 (2015)

\bibitem{Scoppola2015}
 B. Scoppola, C. Lancia e R. Mariani, arXiv:1409.0268

\bibitem{bib:L} T.M. Liggett:
 {\it Stochastic Interacting Systems:
    Contact, Voter and Exclusion Processes}, Springer, New York (1999)

\bibitem{bib:Schue}
     G.M. Sch\"utz,
 in {\it Phase Transitions and Critical Phenomena vol 19.}, C. Domb and
 J.~L.~Lebowitz Ed., Academic Press, San Diego  (2001)

\bibitem{bib:ZS}
R. K. P. Zia and B. Schmittmann,
 J. Stat. Mech. (2007) P07012

\bibitem{bib:SCN}
A. Schadschneider, D. Chowdhury, K. Nishinari:
{\it Stochastic Transport in Complex Systems: From Molecules to Vehicles},
Elsevier Science, Amsterdam (2010)

\bibitem{bib:KRB}
P.L. Krapivsky, S. Redner, E. Ben-Naim:
{\it  A Kinetic View of Statistical Physics},
Cambridge University Press, Cambridge (2010)

\bibitem{bib:BE}
R.A. Blythe, M.R. Evans: J. Phys. A: Math. Gen. 40, R333 (2007)

\bibitem{bib:D} B. Derrida: J. Stat. Mech. (2007) P07023

\bibitem{Schuetz1} G. Sch\"utz: J. Stat. Phys. 71, 471 (1993)

\bibitem{Hinrichsen} H. Hinrichsen, S. Sandow: J. Phys. A 30, 2745 (1997)

\bibitem{kolomeiski2}
A.B. Kolomeisky:
J. Phys. A 31, 1153 (1998)


\bibitem{barma2}
G. Tripathy, M. Barma:
Phys. Rev. E {\bf 58}, 1911 (1997)


\bibitem{chou}
T.\ Chou, G.W.\ Lakatos:
Phys. Rev. Lett. 93, 198101 (2004)

\bibitem{LakatosBC}
G.W.\ Lakatos, J. O'Brien, T. Chou: J. Phys. A 39, 2253 (2006)

\bibitem{harris1}
R.J. Harris, R.B. Stinchcombe:
Phys. Rev. E {\bf 70}, 016108 (2004)

\bibitem{Enaud}
C. Enaud, B. Derrida:
Europhys. Lett. {\bf 66}, 83 (2004)

\bibitem{juhasz}
R. Juhasz, L. Santen, F. Igloi:
Phys. Rev. E {\bf 74}, 061101 (2006)

\bibitem{frey2}
P. Pierobon, M. Mobilia, R. Kouyos, E. Frey:
Phys. Rev. E 74, 031906 (2006)

\bibitem{dong}
J.J. Dong, B. Schmittmann, R.K.P. Zia:
J. Stat. Phys. 128, 21 (2007)

\bibitem{foulaad}
M.E. Foulaadvand, S. Chaaboki, M. Saalehi:
Phys. Rev. E {\bf 75}, 011127 (2007)

\bibitem{GreulichS1}
P. Greulich, A. Schadschneider: Physica A387, 1972 (2008)

\bibitem{GreulichS2}
P. Greulich, A. Schadschneider: J. Stat. Mech. (2008) P04009

\bibitem{GreulichS3}
P. Greulich, A. Schadschneider: Phys. Rev. E 79, 031107 (2009)

\bibitem{MCupdate} A single Monte Carlo update consists in choosing a
  bond at random, and updating its configuration according to
  dynamical rules.

\bibitem{DerridaEvans93}
B. Derrida, M. Evans: J. Physique I 3, 311 (1993)

\bibitem{Random_TEST}
P. L'Ecuyer, R. Simard: ACM Trans. Math. Softw. 33, 22 (2007)

\bibitem{Park_Miller}
S. K. Park, K. W. Miller: Commun. ACM 31, 1192 (1988)



\bibitem{Embley}
B. Embley, A. Parmeggiani, N. Kern: Phys. Rev. E 80, 041128 (2009)

\bibitem{Ezaki}
T. Ezaki, K. Nishinari: J. Stat. Mech. (2012) P11002

\bibitem{Raguin}
A. Raguin, A. Parmeggiani, N. Kern: Phys. Rev. E 88, 042104 (2013)

\bibitem{Neri}
I. Neri, N. Kern, A. Parmeggiani: New J. Phys. 15, 085005 (2013)

\bibitem{Meesoon-recent}
Y. Baek, M. Ha, J. Jeong: Phys. Rev. E 90, 062111 (2014)

\bibitem{Diedrich}
G. Diedrich, L. Santen, A. Schadschneider, J. Zittartz:
Int. J. Mod. Phys. C 11, 335 (2000)

\end{thebibliography}

\end{document}